\documentclass[prl,epsfig,twocolumn,showpacs,preprintnumbers,amsmath,amssymb]{revtex4}
\usepackage{amsmath}
\usepackage[usenames]{color}
\usepackage{epsfig}
\usepackage{graphics}
\begin{document}

\title{Antiferromagnetic noise correlations in optical lattices}
\author{G.\ M.\ Bruun$^{1,2}$, O.\ F.\ Sylju{\aa}sen$^3$, K.\ G.\ L.\ Pedersen$^4$, B.\ M.\ Andersen$^4$, E.\ Demler$^5$, A.\ S.\ S\o rensen$^4$} 
\affiliation{$^1$Niels Bohr International Academy, University of Copenhagen,  DK-2100 Copenhagen \O, Denmark\\
$^2$Mathematical Physics, Lund Institute of Technology, P.\ O.\ Box 118, SE-22100 Lund, Sweden\\
$^3$Department of Physics, University of Oslo, P.~O.~Box 1048 Blindern, N-0316 Oslo, Norway\\
$^4$Niels Bohr Institute, University of Copenhagen, 
DK-2100 Copenhagen \O, Denmark\\
$^5$Department of Physics, Harvard University, Cambridge, Massachusetts 02138, USA}
\date{\today{}}

\begin{abstract}
We analyze how noise correlations probed by time-of-flight (TOF) 
experiments reveal antiferromagnetic (AF) correlations of fermionic atoms in two-dimensional (2D) and 
three-dimensional (3D) optical lattices. Combining analytical and quantum Monte Carlo (QMC) calculations using
 experimentally realistic parameters, we show that AF correlations can be detected
  for temperatures above and below the critical temperature for AF ordering. It is demonstrated that
spin-resolved noise correlations yield important information about the spin ordering. Finally, we show 
how to extract the spin correlation length and the related critical exponent of the AF transition from the noise. 
\end{abstract}
\maketitle

Atoms in optical lattices hold the potential to unravel the fundamental physics of phenomena related to 
 quantum systems in periodic potentials including spin phases and high-$T_c$ superconductors~\cite{Hofstetter}. One has already 
observed the Mott insulator transition for bosons~\cite{Greiner1}, the emergence of superexchange interactions~\cite{Trotzky}, the 
transition between metallic, band-insulator and Mott phases for fermions~\cite{Jordens,Schneider}, and fermionic pairing~\cite{Chin}.
In addition to creating these lattice systems at sufficiently low temperatures $T$, a  major challenge is how to detect the various  phases
predicted theoretically. These phases can be investigated via higher order correlation functions which only show up as
 quantum noise in most experiments. Quantum spin noise spectroscopy~\cite{Eckert,Bruun}  and 
 the measurement of correlations of the momentum distribution of the atoms 
after release from the lattice (TOF experiments) are two ways to probe these correlation functions~\cite{Altman}. 
TOF experiments have already been used to detect pairing correlations in a Fermi gas~\cite{Greiner2}, 
bosonic bunching and fermionic anti-bunching of atoms in optical lattices~\cite{Folling},  the Mott-superfluid transition for bosons~\cite{Spielman}, and 
the effects of disorder in the Mott phase~\cite{Guarrera}.

\begin{figure}
\includegraphics[width=0.9\columnwidth,height=0.6\columnwidth]{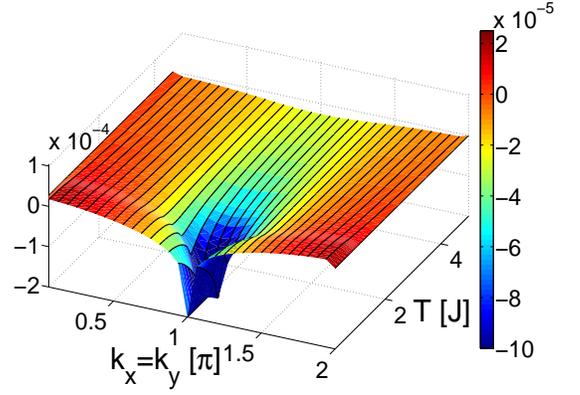}
\caption{(Color online) Plot of $C^{\rm{exp}}_{\rm nn}({\mathbf k}_\perp)$ versus temperature $T$ and momentum along a diagonal cut $k=k_x=k_y$. 
}
\label{pipiover43Dfig}
\end{figure}

\begin{figure}
\includegraphics[clip=true,height=0.47\columnwidth,width=0.9\columnwidth]{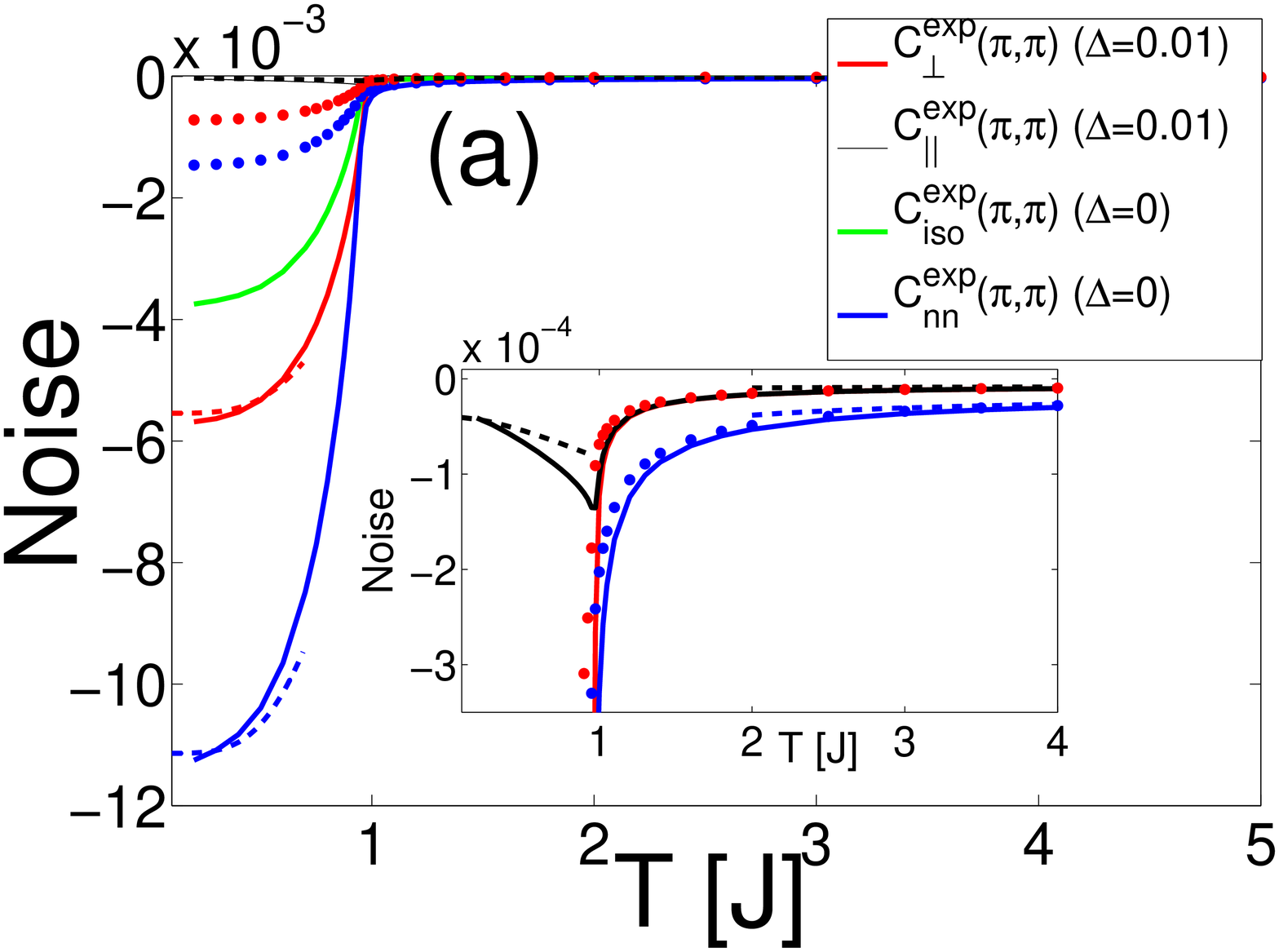}
\includegraphics[clip=true,height=0.47\columnwidth,width=0.9\columnwidth]{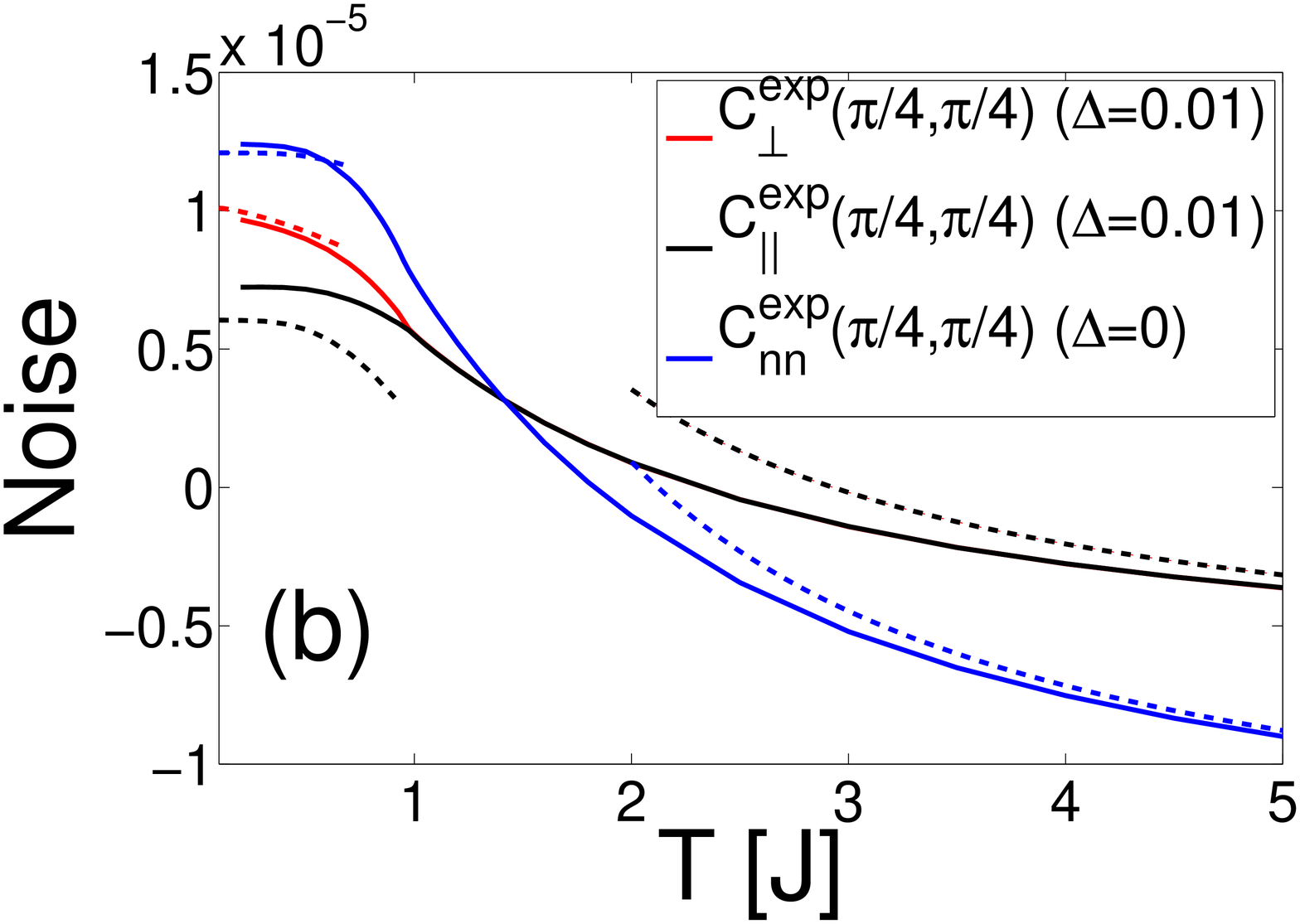}
\caption{(Color online) Noise correlation functions $C^{\rm{exp}}_{\perp}({\mathbf k}_\perp)$ (red), $C^{\rm{exp}}_{\parallel}({\mathbf k}_\perp)$ (black), $C^{\rm{exp}}_{\rm iso}({\mathbf k}_\perp)$ (green), and $C^{\rm{exp}}_{\rm nn}({\mathbf k}_\perp)$ (blue) versus $T$ at ${\mathbf k}_\perp =(\pi,\pi)$ (a) 
and ${\mathbf k}_\perp =(\pi/4,\pi/4)$ (b). The dotted (solid) lines are with (without) Gaussian smearing ($\kappa=1/40$). Dashed lines show the 
analytical results   (\ref{CnnhighT})-(\ref{CudlowT}). Inset (a): same plot but zoomed in near $T_N$. }
\label{pipi3Dfig}
\end{figure}

Here, we show how  AF correlations of fermionic atoms in  optical lattices give rise to  distinct measurable signals 
in TOF experiments even above  the critical temperature for magnetic ordering. Spin-resolved  experiments are 
demonstrated to yield additional information which can  be used to identify the magnetic ordering and broken symmetry axis. 
The main results are illustrated in Figs.~\ref{pipiover43Dfig}-\ref{pipi3Dfig}
 which show noise correlations in the momentum distributions after expansion as 
a function of temperature and momentum. We finally discuss how the spin correlation length and 
 the related critical exponent $\nu$ can be extracted experimentally from the noise. 

We consider a two-component Fermi gas in an optical lattice of size $N=N_xN_yN_z$. 
In the limit of strong repulsion the gas is in the Mott phase for low $T$ at half-filling and can be described by the Heisenberg model
\begin{equation}
\hat H=J\sum_{\langle l,m\rangle}[\hat s^x_l\hat s^x_m+\hat s^y_l\hat s^y_m+(1+\Delta)\hat s^z_l\hat s^z_m].
\label{Heisenberg}
\end{equation} 
Here $\hat{\mathbf s}_l$ is the spin-$1/2$ operator for atoms at site ${\mathbf r}_l$ and $\langle l,m\rangle$ denotes neighboring pairs. 
The interaction is $J=4t_\uparrow t_\downarrow/U$, 
with $U>0$ the on-site repulsion between atoms and $t_\sigma$
the spin-dependent tunneling matrix element. The anisotropy parameter is 
$\Delta\!=-(t_\uparrow-t_\downarrow)^2/(t_\uparrow^2+t_\downarrow^2)$. 
Below we consider both cubic (3D) and square (2D) lattices with lattice constant $a$ of unity. We do not include any effects of a trapping potential. 

A major experimental goal is to detect the onset of AF correlations with decreasing $T$.
 TOF experiments probe  correlation functions of the form~\cite{Folling}   
\begin{equation}
C_{AB}({\mathbf{r}}-{\mathbf{r}}')=\frac{\langle \hat A({\mathbf{r}})\hat B({\mathbf{r}}')\rangle-\langle \hat A({\mathbf{r}})\rangle\langle \hat B({\mathbf{r}}')\rangle}{\langle \hat A({\mathbf{r}})\rangle\langle \hat B({\mathbf{r}}')\rangle},
\label{Cgeneral}
\end{equation}
where $\hat A({\mathbf r})$ and $\hat B({\mathbf r})$ are atomic observables measured at ${\mathbf r}$ after expansion. 
In Refs.~\onlinecite{Greiner2,Folling}, 
the atomic density correlations $C_{\rm nn}({\mathbf{d}})$ were measured 
with $\hat A({\mathbf r})=\hat B({\mathbf r})=\sum_\sigma\hat n_\sigma({\mathbf{r}})$ the density operator 
of the atoms. Here $\hat n_\sigma({\mathbf{r}})=\hat\psi_\sigma^\dagger({\mathbf{r}})\hat\psi_\sigma({\mathbf{r}})$ 
with $\hat\psi_\sigma({\mathbf{r}})$ the field operator for  atoms in spin state $\sigma$. We also consider the correlations between  images of each of the  spin components $C_{\parallel}({\mathbf{d}})$ corresponding to $\hat A({\mathbf r})=\hat n_\uparrow({\mathbf{r}})$ and  $\hat B({\mathbf r})=\hat n_\downarrow({\mathbf{r}})$. This may be achieved by, e.g., spatially separating the two atomic spin states using a 
Stern-Gerlach technique~\cite{Esslinger}. By applying a $\pi/2$ pulse before the expansion one can also gain access to the spin noise perpendicular to the $z$ component $C_{\perp}({\mathbf{d}})$ corresponding to $\hat A({\mathbf r})=\exp(i\hat s^y\pi/2)\hat n_\uparrow({\mathbf{r}})\exp(-i\hat s^y\pi/2)$ and  $\hat B({\mathbf r})=\exp(i\hat s^y\pi/2)\hat n_\downarrow({\mathbf{r}})\exp(-i\hat s^y\pi/2)$.

 After normal ordering and expansion of the field operators in the lowest band Wannier states, we obtain 
\begin{eqnarray}
C_{\rm nn}({\mathbf{k}})&=&\frac 1 N -\frac 1 2 \delta_{{\mathbf k},{\mathbf K}}-2\langle\hat{\mathbf{s}}_{\mathbf k}\cdot\hat{\mathbf{s}}_{-{\mathbf k}}\rangle,
\label{Cnn}\\
C_{\parallel}({\mathbf{k}})&=&\frac 1 {2N}-\langle \hat s^z_0\hat s^z_0\rangle-\langle \hat s^x_{\mathbf k}\hat s^x_{-{\mathbf k}}\rangle
-\langle \hat s^y_{\mathbf k}\hat s^y_{-{\mathbf k}}\rangle,
\label{Cud}\\
C_{\perp}({\mathbf{k}})&=&\frac 1 {2N}-\langle \hat s^x_0\hat s^x_0\rangle-\langle \hat s^y_{\mathbf k}\hat s^y_{-{\mathbf k}}\rangle
-\langle \hat s^z_{\mathbf k}\hat s^z_{-{\mathbf k}}\rangle,
\label{Clr}
\end{eqnarray}
where  ${\mathbf K}$ is a reciprocal lattice vector, and  $\hat{\mathbf{s}}_{\mathbf k}=N^{-1}\sum_l\hat{\mathbf s}_le^{-i{\mathbf k}\cdot{\mathbf r}_l}$.
We  assume free expansion of the atoms for a duration $t$,
 neglect autocorrelation terms $\propto \delta({\mathbf{d}})$ in (\ref{Cnn})-(\ref{Clr}), and express $C$ in terms of the 
  momentum ${\bf k}=m{\bf d}/t$ ($\hbar=1$ throughout).
 These correlation functions have contributions with  different scalings. 
 For 3D systems, the spins order below the N\'eel temperaure $T_N$. 
 Assuming the broken symmetry axis along the $z$ axis, $C_{{\rm nn}}$ and $C_\perp$ have a contribution from $\langle \hat s^z_{{\bf k}} \hat s^z_{{\bf k}}\rangle\sim{\mathcal O}(1)$ at ${\bf k}=(\pi,\pi,\pi)$ for $T<T_N$. 
  At other momenta or for $T>T_N$ the correlation functions scale as $1/N$.
Note that we will perform calculations where the total spin $\propto \hat s_{0}^z$ of the lattice is allowed to fluctuate, 
which is different from the experimental situation where the number of particles in each spin state is fixed. 
 This may give rise to different momentum independent terms in (\ref{Cnn})-(\ref{Clr}) for the experiment,
  but the $k$ dependent part should be accurately captured by our calculations.

Experiments take 2D column density images of the expanding cloud
 corresponding to integrating over $z$ and $z'$ in both the numerator and denominator in (\ref{Cgeneral}). 
The cameras also introduce a smoothening in 
the $xy$-plane, which can be modeled by convolution with a Gaussian~\cite{Folling}. 
In total, the experimental procedure corresponds to measuring the averaged correlation function 
\begin{equation}
C^{\rm{exp}}({\mathbf{k}}_\perp)=
\frac{ 1}{4\pi N\kappa^2}
\sum_{{\mathbf k}'}
e^{-\frac{1}{4\kappa^2}\frac{1}{(2\pi)^2}({\mathbf{k}}'_\perp-{\mathbf{k}}_\perp)^2}
C({\mathbf{k}}'),
\label{ExpCorr}
\end{equation}
where ${\mathbf{k}}_\perp=(k_x,k_y)$ and $\kappa=w/l$ with $l=2\pi t/ma$   (keeping $a$ for clarity) and $w$ a width depending 
 on the pixel resolution  of the CCD camera.  
The averaging in the $z$-direction reduces the contributions to $C_{{\rm nn}}$ and $C_\perp$ at ${\bf k}=(\pi,\pi,\pi)$  from $\mathcal{O}(1)$ for $T<T_N$ to $1/N_z$ and the Gaussian smoothening further reduces it to ${\mathcal O}(1/\kappa^2 N)$. This reduction happens because the correlations are restricted to a single point ${\bf k}=(\pi,\pi,\pi)$ 
 and we are averaging over $~N_z \cdot N_xN_y/\kappa^2$ points.
 Away from ${\mathbf{k}}_\perp=(\pi,\pi)$ or for $T>T_N$ the correlations have a wider distribution and they are less affected by the averaging. 

 We plot in Fig.\ \ref{pipiover43Dfig} $C^{\rm{exp}}_{\rm nn}(k,k)$ as a function of $k=k_x=k_y$ and $T$
for a 3D lattice of size $N=32^3$. The spin correlation functions $\langle\hat{\mathbf{s}}_{\mathbf k}\cdot\hat{\mathbf{s}}_{-{\mathbf k}}\rangle$
were calculated using QMC simulations using the Stochastic Series Expansion method~\cite{SSE} 
 with directed-loop updates~\cite{SS}. This method is very efficient for Heisenberg models and gives accurate results for a wide range of $T$ 
 for large systems. For the isotropic Heisenberg model,  the QMC calculations 
   yield $T_N\simeq 0.945J$ in agreement with Ref.\ \onlinecite{Sandvik}. 
  $C^{\rm{exp}}_{\rm nn}$ was then calculated from (\ref{Cnn}) with the $k_z$ average in 
 (\ref{ExpCorr}) included to simulate the experimentally relevant situation~\cite{Folling}. 
The main feature of the plot is the dip at $k=\pi$ coming from AF ordering. For  $T<T_N$, this gives rise to a large Bragg dip
at $k_x=k_y=\pi$~\cite{Altman,Andersen}. 
We see that the Bragg dip remains also above $T_N$ but has a larger width 
due to AF correlations  without long range order. The spin correlation length $\xi$ can be extracted 
directly from the width of the dip which decreases with decreasing $T-T_N$ reflecting that $\xi$ increases as $\xi\sim 1/|T-T_N|^\nu$ for $T\rightarrow T_N^+$.
We show below how the noise  indeed can be used to extract the critical exponent $\nu$. 
At high $T$ the AF correlations lead to singlet formation of neighboring spins. This gives rise to a precursor of the Bragg peak at $k_x=k_y=\pi$
and an equal signal of opposite sign at ${\mathbf k}_\perp=0$ [see (\ref{CnnhighT})].
These momentum correlations can be understood by noting that two fermions in a singlet are more (less) likely to have the 
same (opposite) momentum due to the Pauli exclusion principle. 
The  uniform spin noise case $k=0$ was also considered in~\cite{Bruun}.

In the high-$T$ limit $J/T=1/\tilde T\ll 1$ ($k_B=1$), controlled analytical results for
 the correlation functions (\ref{Cnn})-(\ref{Clr}) may be obtained by  expanding  in $\tilde T^{-1}$.  We obtain   
\begin{gather}
C_{\rm nn}({\mathbf{k}})=-\frac 1 2 \delta_{{\mathbf k},{\mathbf K}}-\frac 1 {2N}+  \frac {3Z\gamma_{\mathbf k}} {8N\tilde T}+{\mathcal O}({\tilde T}^{-2}),
\label{CnnhighT}\\
C_{\parallel}({\mathbf{k}})\!=\!C_{\perp}({\mathbf{k}})=-\frac 1 {4N}+\frac{Z}{8N\tilde T}(\frac 1 2 +\gamma_{\mathbf k})+{\mathcal O}({\tilde T}^{-2})
\label{CudhighT}
\end{gather}
 with $Z=4$($6$) for 2D(3D) lattices. For simplicity we have taken $\Delta=0$ and  
$\gamma_{\mathbf k}=Z^{-1}\sum_{\mathbf a}2\cos({\mathbf k}\cdot {\mathbf a})$ where $\sum_{\mathbf a}$ sums over 
the unit vectors spanning the lattice. The average of (\ref{CnnhighT}) and (\ref{CudhighT}) over $z$ can be obtained by simply omitting the $z$ direction in the sum.

To obtain analytic expressions  for low $T$, we perform a 
spin-wave calculation, which  should be fairly accurate for a 3D system at $T\ll T_N$. This yields after some algebra
\begin{eqnarray}
C_{\parallel}({\mathbf{k}})\!=\!\frac 1 {2N}-\langle \hat s^z_0\hat s^z_0\rangle-\frac 1 {2N}(1+2f_{\tilde {\mathbf  k}})e^{-2\Theta_{{\mathbf  k}}},
\label{CudlowT}
\end{eqnarray}
where $\tanh 2\Theta_{\mathbf k}=\gamma_{\mathbf k}/(1+\Delta)$ and $f_{\mathbf k}=[\exp\omega_{\mathbf k}/T-1]^{-1}$ is the Bose distribution function
for the spin-waves with energy  $\omega_{\mathbf k}=3J\sqrt{(1+\Delta)^2-\gamma^2_{\mathbf k}}$. 
 We also have $\langle \hat s_0^z\hat s_0^z\rangle={\sum_{{\mathbf k}}}' \sinh^{-2}(\beta\omega_{\mathbf k}/2)/2N^2$
 where the sum extends over the AF reduced Brillouin zone ~\cite{Bruun}.  
  A similar but somewhat more lengthy expression can
 be derived for $C_{\perp}({\mathbf{k}})$.

We plot in Fig.\ \ref{pipi3Dfig} $C^{\rm{exp}}_{\rm nn}({\mathbf k}_\perp)$, $C^{\rm{exp}}_{\parallel}({\mathbf k}_\perp)$,
 and $C^{\rm{exp}}_{\perp}({\mathbf k}_\perp)$ as a function of $T$ for ${\mathbf k}_\perp =(\pi,\pi)$ (a) and ${\mathbf k}_\perp =(\pi/4,\pi/4)$ (b)
 for the same parameters as above.   The solid curves include the average over $k_z$  only and the dotted curves include a Gaussian 
smoothing in the $(k_x,k_y)$ plane as well. The values for $T\rightarrow 0$
 with Gaussian smearing are consistent with the results of Refs.~\cite{Folling} given the different system size and slightly different measured quantities.
For simplicity we have excluded the Gaussian averaging for ${\mathbf k}_\perp =(\pi/4,\pi/4)$
since it [contrary to the ${\mathbf k}_\perp =(\pi,\pi)$ case]  leads to only negligible changes.
We trigger the AF ordering along the $z$-direction for $T<T_N$ by
including a small anisotropy $\Delta=0.01$ for $C^{\rm{exp}}_{\parallel}$ and $C^{\rm{exp}}_{\perp}$. 

Figure \ref{pipi3Dfig}(a) shows the  Bragg dip for $T<T_N$ coming  from 
 $\langle \hat s^z_{{\bf k}} s^z_{{\bf -k}}\rangle \approx |\langle s^z_i\rangle|^2$  for ${\bf k}= (\pi,\pi,\pi)$ with $|\langle s^z_i\rangle|>0$. 
Extrapolating to $T=0$ we have $ |\langle s^z_i\rangle|\simeq 0.43 \pm 0.01$,  a value reduced from $1/2$ by quantum  fluctuations.
On the other hand the spin noise parallel to the broken 
symmetry axis  $C_\parallel$ is reduced with the onset of magnetic ordering for $T<T_N$ as can be seen from the inset in Fig.\ \ref{pipi3Dfig}(a).
 (The minimal value of $C_\parallel$ at $T_N$ is  dependent on the anisotropy $\Delta$.)  Note that even though the 
 density correlations are larger and hence more easily measurable, the spin resolved measurements 
are crucial in verifying e.g. whether the correlation dip is indeed due to
magnetic ordering and not caused, for example, by formation of a period-doubled charge-density wave. 
The difference between $C_\parallel$ and $C_\perp$ for $\Delta>0$ can furthermore be used to identify the 
broken symmetry axis. In the isotropic case, there is no broken symmetry axis and  
$C^{\rm{exp}}_\parallel=C^{\rm{exp}}_\perp=C_{\rm iso}^{\rm{exp}}$  for $\Delta\rightarrow 0$,
  as can be seen from the green curve in Fig.\ \ref{pipi3Dfig}(a).

The calculated correlations are rather small even for the quantities including the macroscopic contribution $\langle \hat s^z_{{\bf k}} s^z_{{\bf -k}}\rangle \approx |\langle s^z_i\rangle|^2$  at ${\bf k}= (\pi,\pi,\pi)$, 
e.g., for low temperatures  $T\lesssim J/2$ $C_{{\rm nn}}^{\rm{exp}}\approx10^{-2}$ ($10^{-3}$) without (with) Gaussian smoothening and scales like $1/N_z$ ($1/N$). This is, however, still significantly larger than the correlations of order $C_{{\rm nn}}\sim 10^{-4}$ which were recently measured with slightly bigger system sizes~\cite{Folling} . The correlations are also close to the measured experimental values even for temperatures above $T_N$, i.e., for $T=2 J$ we have $C^{\rm{exp}}_{{\rm nn}}\approx 6\cdot  10^{-5}$  (scaling as $1/N$), which is comparable to the recent experiments \cite{Folling}.  The measurement of noise correlations can thus be used to show the onset of AF order even above the critical temperature. Furthermore, 
for current experiments the spin temperature is uncertain because there are no sensitive probes. If the noise correlations are measured, the detailed theoretical curves presented here would provide a means of assessing the spin temperature in the experiments. 

The density and spin noise at ${\mathbf k}_\perp =(\pi/4,\pi/4)$ depicted in Fig.\ \ref{pipi3Dfig}(b)
exhibit  a different behavior from that at ${\mathbf k}_\perp =(\pi,\pi)$:
it now  decreases in numerical value for high $T$ and even  changes sign above $T_N$. 
 This unusual behavior is a geometric effect of the lattice.   Note that since the noise scales as $1/N$ away from the Bragg point, a measurement 
  requires higher experimental resolution than what is presently available
   or more sensitive detection  methods such as spin noise spectroscopy~\cite{Bruun}.
\begin{figure}
\begin{center}
\leavevmode
\begin{minipage}{.45\columnwidth}
\includegraphics[clip=true,height=1.0\columnwidth,width=1\columnwidth]{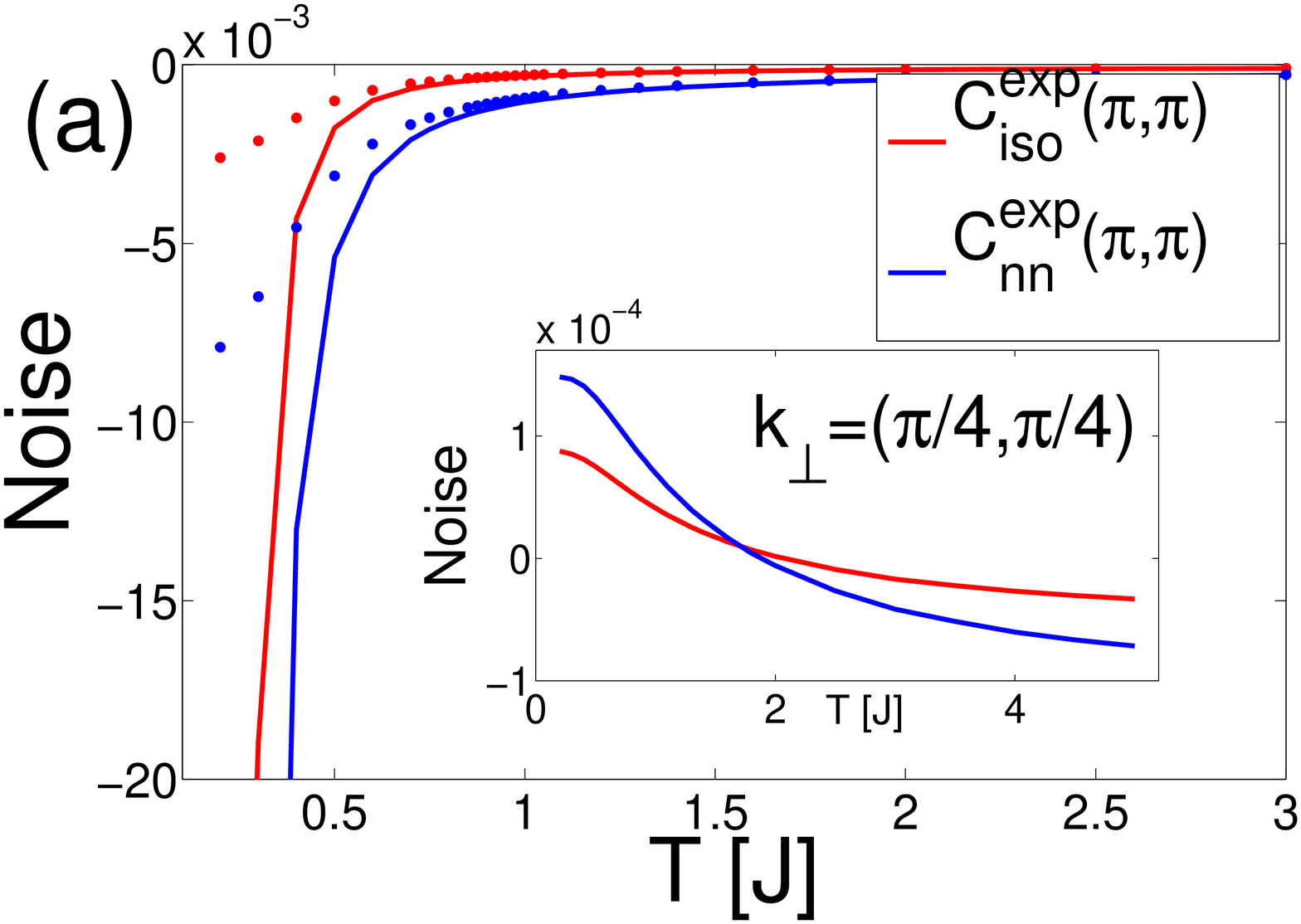}
\end{minipage}
\begin{minipage}{.53\columnwidth}
\includegraphics[clip=true,height=1.0\columnwidth,width=1\columnwidth]{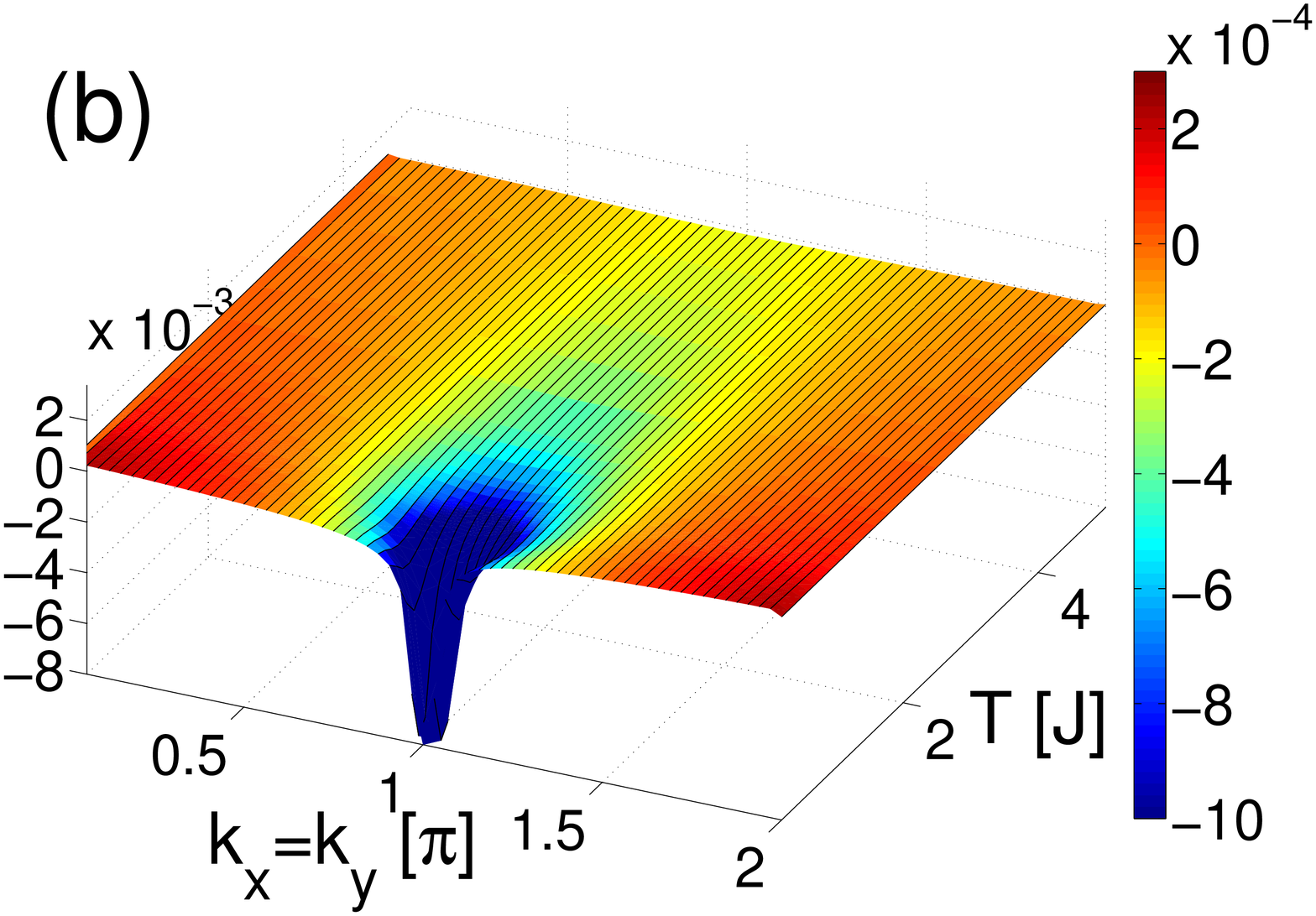}
\end{minipage}
\caption{(Color online) (a) Noise correlation functions $C^{\rm{exp}}_{\rm iso}({\mathbf k}_\perp)$ (red) and $C^{\rm{exp}}_{\rm nn}({\mathbf k}_\perp)$ (blue) versus $T$ at ${\mathbf k}_\perp=(\pi,\pi)$ for a 2D lattice. The curves are obtained from QMC simulations on $64\times64$ lattices. The dotted (solid) lines are with (without) Gaussian smearing using $\kappa=1/40$. Inset: same as solid lines in main panel except at ${\mathbf k}_\perp =(\pi/4,\pi/4)$. 
(b) Same as Fig. \ref{pipiover43Dfig} but for a 2D system without the $k_z$ summation.}
\label{2Dsystemfig}
\end{center}
\end{figure}

\begin{figure}[b]
\begin{minipage}{.49\columnwidth}
\includegraphics[clip=true,width=0.98\columnwidth]{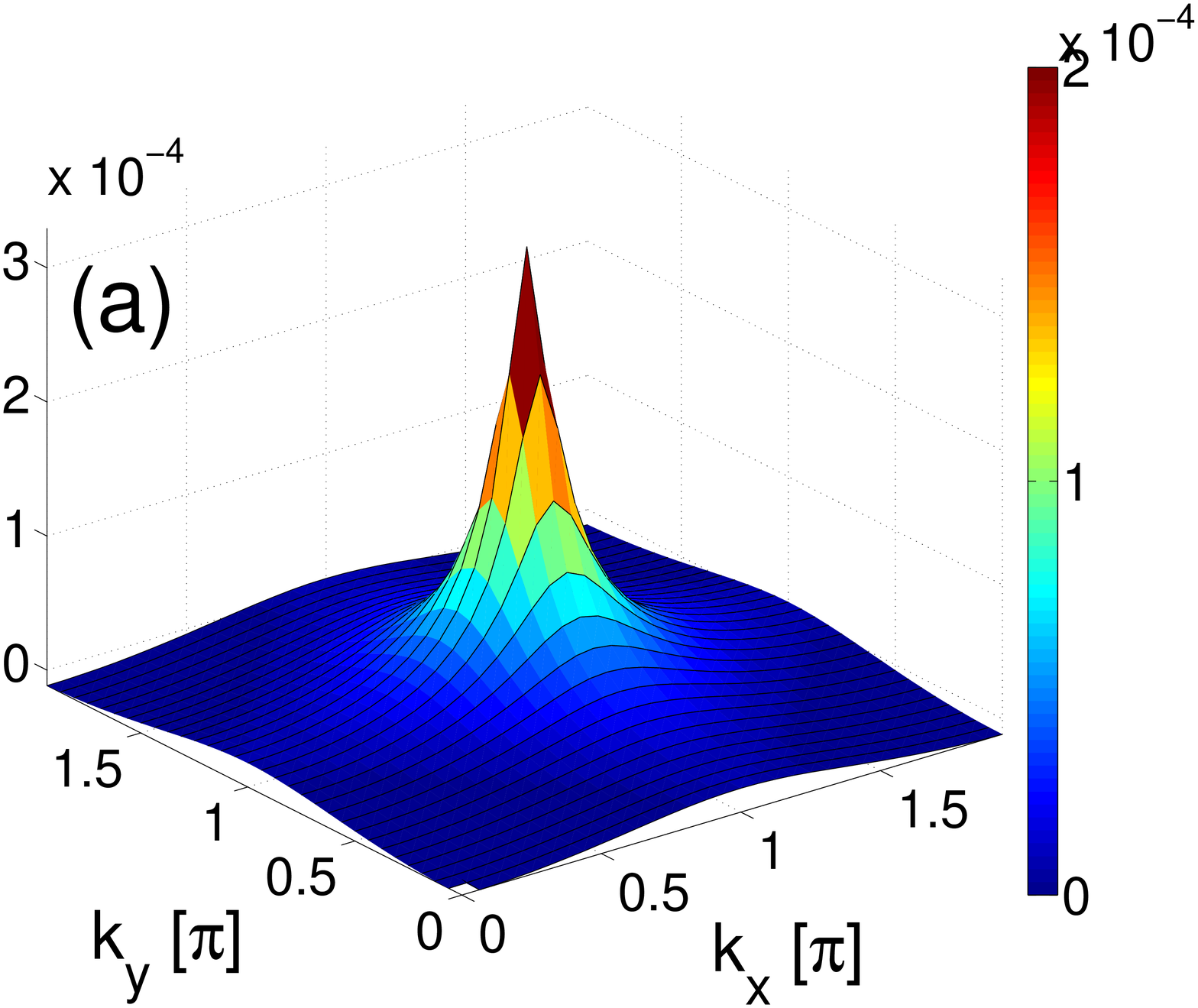}
\end{minipage}
\begin{minipage}{.49\columnwidth}
\includegraphics[clip=true,width=0.98\columnwidth]{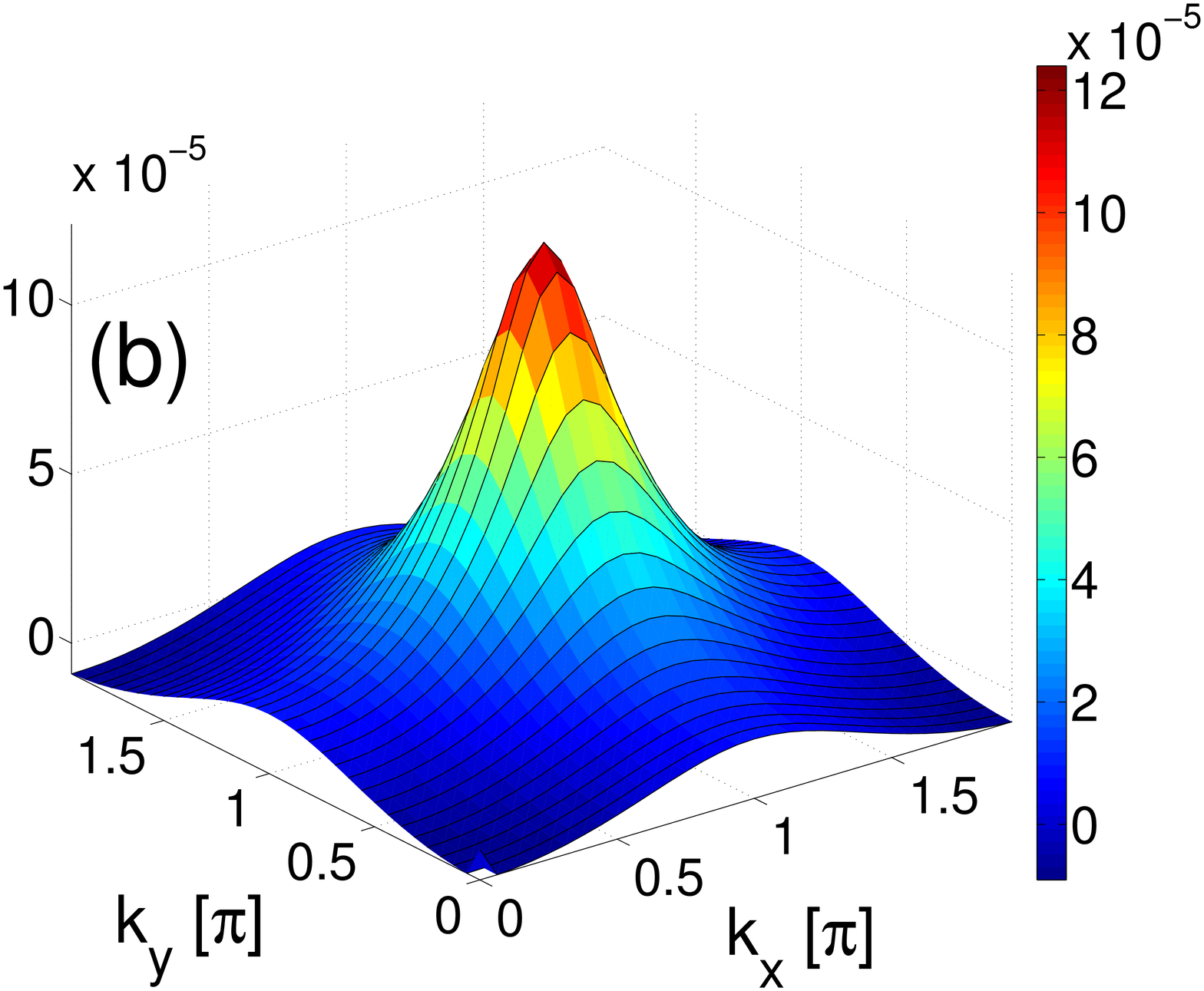}
\end{minipage}
\\
\begin{minipage}{.49\columnwidth}
\includegraphics[clip=true,width=0.98\columnwidth]{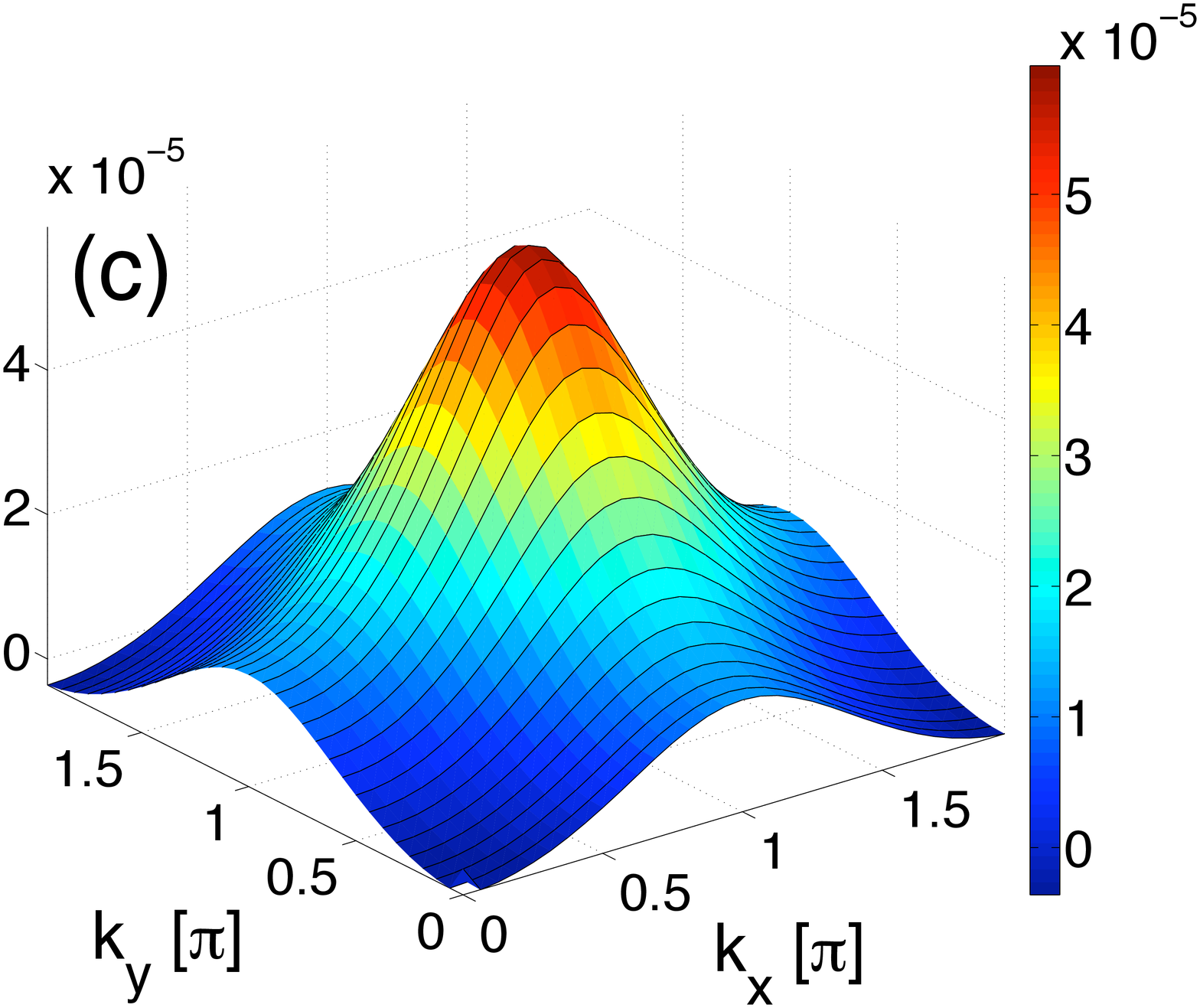}
\end{minipage}
\begin{minipage}{.49\columnwidth}
\includegraphics[clip=true,width=0.98\columnwidth]{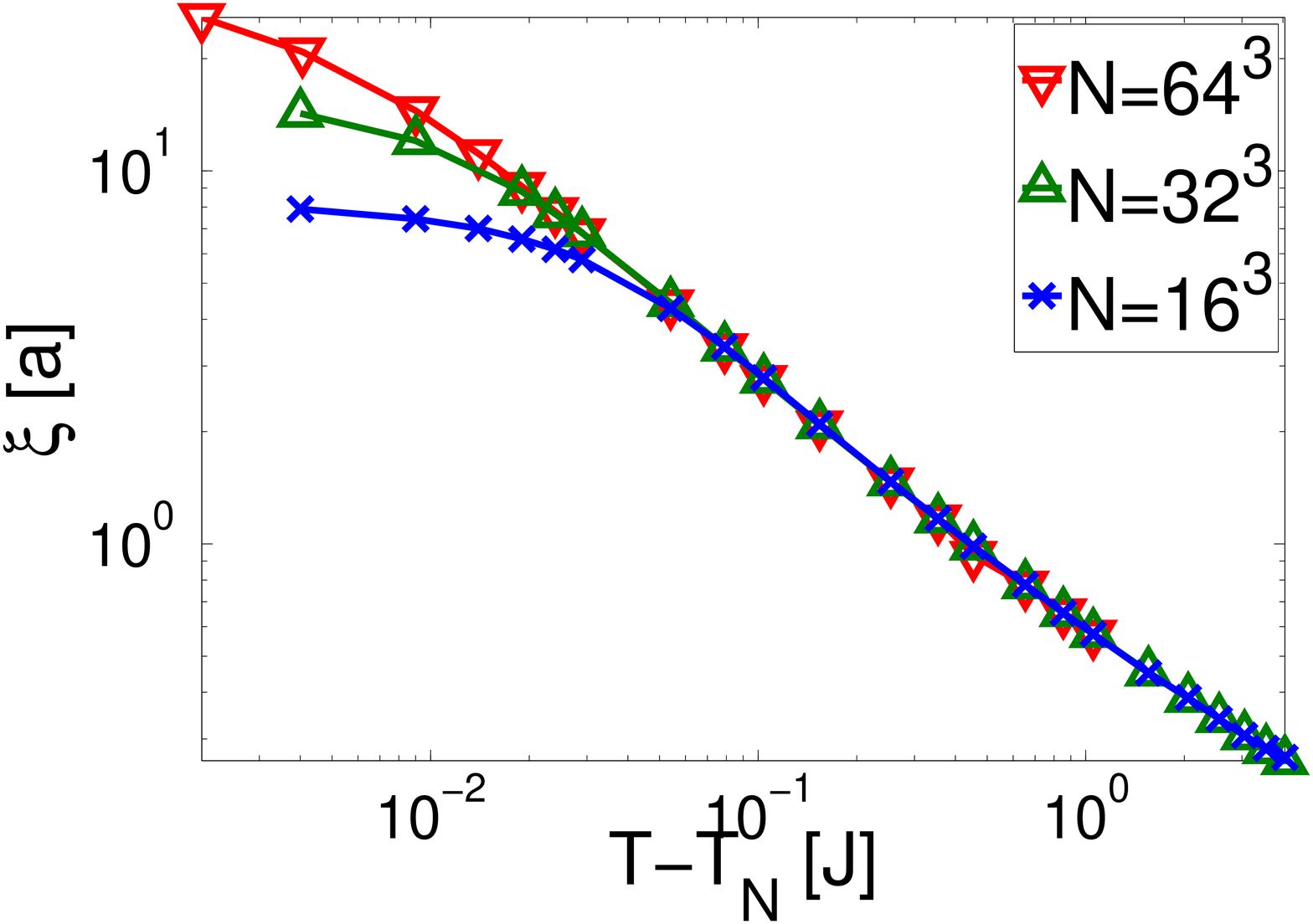}
\end{minipage}
\caption{(Color online) Panels (a-c) show $(-1)C^{\rm{exp}}_{\rm nn}({\mathbf k}_\perp)$ at $T/J=1.0$ (a), $T/J=1.2$ (b), and $T/J=1.6$ (c). (d) Log-log plot of extracted correlation length $\xi$ versus $|T-T_N|$ displaying the power-law behavior $\xi \sim |T-T_N|^{-0.70}$.
} \label{fig4}
\end{figure}

Atomic gases are well suited to study fundamental 
problems in 2D physics as the observation of the Berezinskii-Kosterlitz-Thouless transition  illustrates~\cite{Hadzibabic}.
Recently, single layer 2D atomic gases have been produced which avoids the averaging over $z$ discussed above~\cite{Gemelke}. 
We now study the Mott phase at half-filling for a 2D square lattice. 
In 2D there is no  ordered phase for $T>0$ due to fluctuations, 
but there are still significant AF correlations~\cite{Auerbach}. This is illustrated in Fig.\ \ref{2Dsystemfig}(a), 
which shows $C^{\rm{exp}}_{\rm iso}({\mathbf k}_\perp)$ 
and $C^{\rm{exp}}_{\rm nn}({\mathbf k}_\perp)$ as a function of $T$. The AF correlations give rise to a large dip for ${\mathbf k}_\perp =(\pi,\pi)$ 
both in the density and spin noise which is a precursor of the AF ordered state at $T=0$. Since there is no
averaging over the $z$-direction, the correlations are much stronger than for a 3D system above $T_N$.
 Figure \ref{2Dsystemfig} 
 illustrates that the noise exhibits the same non-trivial features as a 
 function of $k$ and $T$ as the 3D case.

Critical exponents characterizing continuous phase transitions are often difficult to measure. We now demonstrate how 
noise measurements 
can be used to obtain the critical exponent  $\nu$.  
 The correlation length $\xi$ can be extracted from the width of the AF dip in the 3D resolved function
  $C_{\rm nn}({\mathbf{k}})$ at $\mathbf{k}\!\!=\!\!(\pi,\pi,\pi)$. We find that this yields a critical exponent $\nu\approx0.70$ as expected for a 3D 
Heisenberg model \cite{pelissetto}. Importantly, $\nu$ can also be extracted from the experimentally relevant $k_z$-summed 
correlation function $C^{\rm{exp}}_{\rm nn}$. 
Figure \ref{fig4}(a-c) show $C^{\rm{exp}}_{\rm nn}({\mathbf k}_\perp)$ at three fixed temperatures above $T_N$. To obtain $\nu$, we 
fit $C^{\rm{exp}}_{\rm nn}({\mathbf k}_\perp)$ to a summed lattice propagator  of the form 
$\sum_{k_z} [2(3-\sum_{\alpha=x,y,z}\cos (k_\alpha-\pi))\xi^2 +1]^{-1}$ with $k_z=2 \pi n_z/N_z$. 
Figure \ref{fig4}(d) shows the extracted correlation length $\xi$ for various system sizes. One clearly sees the finite-size effects setting in at decreasing $T-T_N$.
 The extracted power-law yields  $\nu\approx0.70$. To obtain 
a robust value for $\nu$ one needs to probe 
the noise for temperatures  where it is somewhat smaller 
than what has been measured to date. This would however enable one to probe  the critical properties of the AF transition. 

In summary, we performed analytic and numerical calculations modeling TOF experiments for repulsive fermionic atoms in 
optical lattices using experimentally realistic parameters. 
This demonstrated that such experiments are well suited to detect AF correlations  both below and above the critical temperature. 
 Spin-resolved measurements were shown to yield valuable additional information and we finally discussed how to extract 
 the critical exponent governing the correlation length close to the AF transition from the noise. 

\begin{acknowledgments}
We thank H.\ Moritz for the suggestion to look at TOF experiments. 
 The QMC simulations were carried out using resources provided by the NOTUR project of the Norwegian Research Council. B.~M.~A. acknowledges support from the Villum Kann Rasmussen foundation and  E.~D.\ from NSF DMR-0705472, CUA, DARPA, MURI.
 \end{acknowledgments}

\end{document}